\documentclass[prb,nofootinbib,twocolumn,superscriptaddress]{revtex4} 

\usepackage{graphicx}
\usepackage{dcolumn}
\usepackage{bm}
\usepackage{threeparttable}
\usepackage{times}
\usepackage{mathptmx}
\usepackage{lscape}
\usepackage{natbib}
\usepackage{amsmath}
\usepackage{amssymb}
\usepackage{braket}
\usepackage{comment}
\usepackage{color}

\def\degree{\kern-.2em\r{}\kern-.3em}

\begin{document}


\title{ Analytic Representation of Canonical Average From Fine Structure of Density of States  }

\author{Koretaka Yuge}
\affiliation{
Department of Materials Science and Engineering,  Kyoto University, Sakyo, Kyoto 606-8501, Japan\\
}%

\author{Shono Ohta}
\affiliation{
Department of Materials Science and Engineering,  Kyoto University, Sakyo, Kyoto 606-8501, Japan\\
}%


\begin{abstract}
{ 
Expectation value of dynamical variables in thermodynamically equilibrium state can be typically provided through well-known canonical average.
The average includes tremendous number of possible states considered far beyond practically handled, which makes it difficult to obtain analytic representation of the average to clarify how the expectation value connects with given interaction: i.e., the relationship is generally understood in \textit{phonomenological} manner except for modest, simple models. 
Here we show that the relationship is explicitly clarified for discrete large systems, where the configurational density of states for any single pair correlation is represented in terms of linear combination of Dirac delta function and its derivatives. 
The significant advantage of the present representation is that it can decompose contributions to macroscopic dynamical variables into harmonicity and anharmonicity in terms of their underlying structural degree of freedom, which will lead to find a set of special microscopic state to characterize macroscopic properties in equilibrium state for classical many-body systems. 
  }
\end{abstract}


\maketitle

\section{Introduction}
For classical thermodynamic system, expectation value of dynamical variables in equilibrium state can be typically determined through the so-called canonical average, i.e., summation over all possible microscopic states on phase space (or for crystalline, solids, can be assumed on configuration space).
Since number of possible states considered exponentially increases with increase of system size, direct determination of the expectation values according to the definition is far from practical even for low-dimensional configuration space, alternative approaches including Moetropolis algorism, entropic sampling and Wang-Landau sampling have been proposed to effectively sample important microscopic states.\cite{mc1,mc2,mc3,wl} 

Very recently, we have found that (i) using only geometric information on configuration space for \textit{non-interacting} system, equilibrium properties and their temperature dependence can be well-characterized by information about a few \textit{a priori known} special microscopic states, and (ii) bidirectional understanding of macroscopic-property/microscopic-interaction relationship, i.e., uniqueness and stability of their correspondence, is theoretically clarified.\cite{em1,em2,em3,em4} These are based on the characteristic geometry of configurational density of states (CDOS), especially focus on the slight deviation of practical CDOS from multidimensional gaussian distribution. 
In order to further study the deviation in CDOS, we recently developed a theoretical approach to systematically determine any order of moment of one-dimensional CDOS along any pair correlation, which shows excellent agreement with that in simulated CDOS.\cite{cdos}
While the moments of CDOS have been quantitatively provided, more explicit form of CDOS itself and the resultant representation of thermodynami average has not been addressed, which should be carefully investigated since it is known that conversion from moment to original function (CDOS) sometimes results in numerical unstability. 
In the present study, we first derive explicit expression of the CDOS in terms of the linear combination of Dirac Delta function, which is applied to predict temperature-dependent short-range order (SRO) parameter, compared with that by conventional thermodynamic average. We find the condition for the present expression to vanish physically non-meaning system size dependence of thermodynamica average, leading to better agreement in SRO including lower odd-order moment of CDOS than that using only even-order moment. However, including higher odd-order moment leads to singularity in SRO at higher temperature, which comes from the collapse of  the positive-semidefinite constraint in CDOS. 

\section{Derivation and Applications}
\begin{figure}[h]
\begin{center}
\includegraphics[width=0.95\linewidth]
{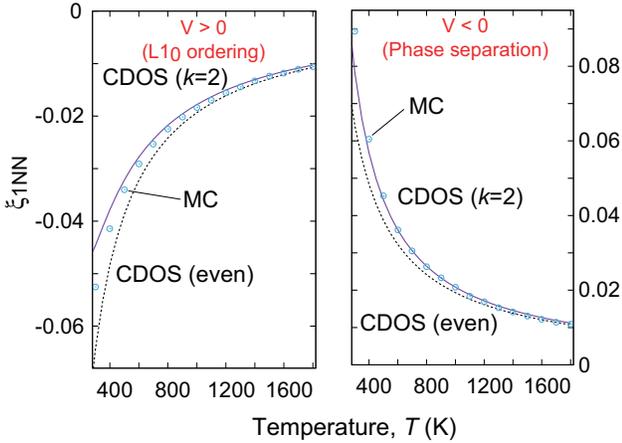}
\caption{Temperature dependence of SRO along nearest-neighbor pair. }
\label{fig:1}
\end{center}
\end{figure}

From our previous study, even- and odd-order moment of CDOS for pair $m$ is given by
\begin{eqnarray}
\label{eq:muf}
\mu_{2\alpha}^{\left(m\right)} &=& \frac{\left( 2\alpha-1\right)!!}{\left(D_mN\right)^\alpha} \nonumber \\
\mu_{2\alpha + 1}^{\left(m\right)} &=& \frac{2M_m\cdot {}_{2\alpha+1}\textrm{C}_3\cdot\left(2\alpha-3\right)!!}{\left(D_mN\right)^{\alpha+1}} ,
\end{eqnarray}
where $D_m$ means number of pair cluster per site, $N$ denotes number of sites, and $M_m$ number of triplets per site, consisting of symmetry-equivalent pair to $m$. 
In order to determine the form of CDOS, $g\left(\xi\right)$ for pair $m$, let us first consider moment-generating function, defined by
\begin{eqnarray}
S\left(t\right) = \int_{-\infty}^{\infty} e^{t\xi} g\left(\xi\right) d\xi = 1+\sum_{n=1}^{\infty} \frac{\mu_n t^n}{t!}. 
\end{eqnarray}
This directly means that $g\left(\xi\right)$ including up to $r$-th order moment can be determined by inverse bilateral Laplace transform of $\mathbb{B}^{-1}$, namely
\begin{eqnarray}
\label{eq:g}
g_r\left(\xi\right) &=& \mathbb{B}^{-1}\left[S\left(-t\right)\right] = \delta\left(\xi\right) + \sum_{k=1}^{r}\left(-1\right)^k \frac{\mu_k}{k!}\delta^{\left(k\right)}\left(\xi\right) \nonumber \\
g\left(\xi\right) &=& \lim_{r\to\infty} g_r\left(\xi\right), 
\end{eqnarray}
where $\delta\left(\xi\right)$ denotes Dirac delta function, and $\delta^{\left(k\right)}\left(\xi\right)$ its $k$-th derivative. 
Substituting Eq.~(\ref{eq:muf}) into Eq.~(\ref{eq:g}), we can obtain explicit expression for $g\left(\xi\right)$ in terms of linear combination of Dirac delta function and its derivatives. 
To perform canonical average of $\xi$ (i.e., SRO), we can simply use the above expression in integral form, namely
\begin{eqnarray}
\label{eq:xi}
\Braket{\xi}\left(T\right) = \dfrac{\displaystyle\int_{-\infty}^\infty \xi\cdot g_r\left(\xi\right) \exp\left(-\beta \Braket{E|\xi} \xi\right) }{\displaystyle\int_{-\infty}^\infty  g_r\left(\xi\right) \exp\left(-\beta \Braket{E|\xi} \xi\right)  },
\end{eqnarray}
where $\Braket{\quad|\quad}$ denotes inner product on configuration space. 
Practical problem to utilize Eq.~(\ref{eq:xi}) is that since $\xi$ is intensive and $\Braket{E|\xi}$ is extensive variable, $\Braket{\xi}\left(T\right)$ explicitly depends on system size $N$ even when $r$ goes to infinity, which is physically non-sence. 
To avoid this, we should increase the maximum order of $r$ by 1 in the denominator of Eq.~(\ref{eq:xi}), namely
\begin{eqnarray}
\Braket{\xi}\left(T\right) = \dfrac{\displaystyle\int_{-\infty}^\infty \xi\cdot g_{r+1}\left(\xi\right) \exp\left(-\beta \Braket{E|\xi} \xi\right) }{\displaystyle\int_{-\infty}^\infty  g_r\left(\xi\right) \exp\left(-\beta \Braket{E|\xi} \xi\right)  }
\end{eqnarray}
Using Eq.~(\ref{eq:muf}), explicit form of the canonical average of SRO including finite-order moment is finally given by
\begin{widetext}
\begin{align}
\label{eq:fin}
\Braket{\xi}\left(T\right)=-\dfrac {1}{\beta\,vN} \dfrac{ \displaystyle\sum _{n=1}^{k+1}{\frac {1}{ \left( 2(n-1)
            \right) !!} \left( {\frac {{\beta}^{2}{v}^{2}N}{{\it D_m}}}
        \right) ^{n}}-{\frac {\beta\,v{\it M_m}}{3{\it D_m}}\displaystyle\sum _{n=1}^{k+1}{\frac {
                \left( 2n+1 \right)}{(2(n-1))!!} \left( {\frac {
                    {\beta}^{2}{v}^{2}N}{{\it D_m}}} \right) ^{n}}} }{\displaystyle\sum _{
        n=0}^{k}{\frac {1}{(2n)!!} \left( {\frac {{\beta}^{2}{v}^{2}N}{{
                    \it D_m}}} \right) ^{n}}-{\frac {\beta\,v{\it M_m}}{3{\it D_m}}\displaystyle\sum _{n=1}^{k}{
            \frac {1}{(2(n-1))!!} \left( {\frac {{\beta}^{2
                    }{v}^{2}N}{{\it D_m}}} \right) ^{n}}} },
\end{align}
\end{widetext}
where $v=\Braket{E|\xi}/N$. 
To demonstrate the validity of Eq.~(\ref{eq:fin}), we estimate temperature dependence of SRO along nearest-neighbor pair on equiatomi fcc lattice, using only even-order moment and using even- and odd-order with $k=1$, compared with conventional thermodynamic average based on Monte Carlo (MC) simulation. 
The result is summarized in Fig.~\ref{fig:1}, where we consider the cases with positive and negative sign of $v$, respectively corresponds to ordering and clustering tendency.  
It is clear from Fig.~\ref{fig:1} that CDOS including odd-order moment leads to better agreement in SRO with that by MC, compared with CDOS including only even-order moment (i.e., this exactly corresponds to gaussian distribution) for a wide range of temperature. At lower temperature, however, we find singularity in SRO for the results with CDOS ($k=2$).
We also find that including higher-order of odd moment leads to the singularity appeared at higher temperature, indicating that higher-order odd moment causes unstable landscape of CDOS. More explicitly, the singularity comes from negative value of partition function (i.e., negative value of CDOS). Since CDOS should be by definition positive semi-definite, this unstability should be due to the missing information lying under the $N$-dependence of moments in Eq.~(\ref{eq:muf}), which has also been pointed out in our previous study. 
In our future work, such missing information would be quantitatively included from microscopic view, which is expected to overcome the problems in $N$-dependence as well as singularity in the canonical average. 

\section{Conclusions}
For classical systems, we provide analytic expression of configurational density of states (CDOS) along any chosen pair on lattice, by employing the analytic expression of its moment based on inverse bilateral Laplace transform. We find that information of the inclusion of odd-order moment should be essential to quantitatively explain temperature dependence of canonical avarage for SRO, while numerical unstability is found when temperature decreases and/or higher-order odd moment is included. This should be quantitatively addressed from microscopic view, which enables to perform canonical average without any trial-and-error simulation.

\section*{Acknowledgement}
This work was supported by a Grant-in-Aid for Scientific Research (16K06704) from the MEXT of Japan, Research Grant from Hitachi Metals$\cdot$Materials Science Foundation, and Advanced Low Carbon Technology Research and Development Program of the Japan Science and Technology Agency (JST).

\end{document}